# Sensitive infrared surface photovoltage in quasi-equilibrium in a layered semiconductor at low-intensity low-temperature condition

Qiang Wan[1§], Keming Zhao[1§], Guohao Dong[2,3§], Enting Li[1], Tianyu Yang[1], Hao Wang[4], Yaobo Huang[5], Yao Wen[4], Yiwei Li[1], Jun He[4,6], Youguo Shi[2,3,7], Hong Ding[8,9], and Nan Xu[1,6*]

*[1] Institute of Advanced Studies, Wuhan University, Wuhan 430072, China*
*[2] Beijing National Laboratory for Condensed Matter Physics and Institute of Physics, Chinese Academy of Sciences, Beijing 100190, China*
*[3] School of Physical Sciences, University of Chinese Academy of Sciences, Beijing 100190, China*
*[4] School of Physics and Technology, Wuhan University, Wuhan 430072, China*
*[5] Shanghai Synchrotron Radiation Facility, Shanghai Advanced Research Institute, Chinese Academy of Sciences, Shanghai 201204, China*
*[6] Wuhan Institute of Quantum Technology, Wuhan 430206, China*
*[7] Songshan Lake Materials Laboratory, Dongguan, Guangdong 523808, China*
*[8] Tsung-Dao Lee Institute & School of Physics and Astronomy, Shanghai Jiao Tong University, Shanghai 200240, China*
*[9] New Cornerstone Science Laboratory, Shanghai 201210, China*

§ These authors contributed equally.
* Email: nxu@whu.edu.cn

**Benefit to layer-dependent bandgap, van der Waals materials with surface photovoltaic effect (SPV) enable photodetection over a tunable wavelength range with low power consumption. However, sensitive SPV in the infrared region, especially in a quasi-steady illumination condition, is still elusive in layered semiconductors. Here, using angle-resolved photoemission spectroscopy, we report a sensitive SPV in quasi-equilibrium in $NbSi_{0.5}Te_2$, with photoresponsivity up to $2.4\times10^6$ V/(W·cm$^{-2}$) at low intensity low temperature condition (LILT). The sensitive SPV is further confirmed by observing the Dember effect, where the photogenerated carrier density is high enough and diffusion currents suppress SPV. Temperature-dependent measurements indicate that intrinsic carriers freezing at low temperature leads to the ultrahigh photoresponse, while a small amount of photon-generated carriers in quasi-equilibrium dominate the system. Our work not only**

**provides a promising layered semiconductor for Infrared optoelectronic devices with strong infrared SPV at LILT, which has application potential in fields such as quantum information and deep-space exploration, but also paves a novel way to enhance light-matter interaction effect by freezing bulk carriers.**

Infrared detection at low-intensity low-temperature (LILT) condition not only provides an invaluable tool for exploring microscopic mechanisms of quantum physics, chemistry and life sciences, but also gives rise to technologies in fields such as quantum information and deep-space exploration [1-4]. Surface photovoltaic effect (SPV) is a typical nonequilibrium phenomenon in semiconductors, induced by photoexcited carriers balancing intrinsic surface band bending [5-7]. Materials with SPV have been widely used in photodetection [8,9], with the advantage of low energy consumption, which makes them suitable in fields such as deep-space exploration. Van der Waals (vdW) semiconductors have drawn substantial attentions in both fundamental research and industrial applications, due to their layer-dependent electrical, optical and magnetic properties [10-13]. Exfoliation of van der Waals materials provides a powerful way to modulate bandgap, which allows for photodetection via the SPV over a broad wavelength range, and makes them key components in next-generation nanodevices [14,15].

Transition metal dichalcogenides (TMDs), one of the most studied 2D semiconductor family, possess a wide bandgap range of 1.0-2.5 eV [16], with SPV beyond the infrared region. Black phosphorus has a narrow bandgap [17,18], however, SPV is only observed within the picosecond time scale by ultrafast experiments [19,20]. Therefore, layered semiconductors with sensitive SPV in the infrared region, especially in quasi-steady low-intensity illumination conditions, are highly desired.

Here, by using angle-resolved photoemission spectroscopy (ARPES), we directly observe a highly sensitive SPV in a layered narrow gap semiconductor $NbSi_{0.5}Te_2$ in a quasi-equilibrium state. The sensitivity of SPV reaches a remarkable value of $2.4\times10^6$ V/(W·cm$^{-2}$) at LILT. With infrared illumination exceeding $10^{-3}$ W/cm$^2$, we observe gradual suppression of SPV due to the Dember effect, where the photogenerated carrier density in quasi-equilibrium is large enough and the diffusion current partially cancels SPV. By performing temperature-dependent measurements, we reveal that the ultrahigh photoresponse is induced by the bulk carriers freezing at low temperature, where photoexcited carriers dominate the system. Our work offers a layered semiconductor with significant promise in infrared optoelectronic device applications in LILT conditions, and reveals a notable way to enhance the sensitivity and efficiency of SPV by carrier freezing.

$NbSi_xTe_2$ is a TMD family with Si-chains inserted in the Nb-plane (Fig. 1a). The Si stoichiometry can be continuously tuned, with one-dimensional massless Dirac fermion [21] and topological hourglass semimetal inherent weak topological insulator [22-24] realized in x = 0.45 and 1/3 compounds, respectively. For the highest Si stoichiometry, $NbSi_{0.5}Te_2$ exhibits a narrow bandgap of 0.4 eV [25-29]. The temperature-dependent Hall results in Fig. 1b indicate a p-type semiconducting behavior of our sample, with hole carrier density dramatically decreasing at lower temperatures (Fig. 1c).

We perform temperature-dependent ARPES measurements to determine band structure evolution near the surface region. The chemical potential deviation between ARPES and Hall measurements demonstrates a surface band bending, which is a precondition for SPV. As shown in Figs. 1e,g, valence bands move downwards as temperature decreases from room temperature. A Lifshitz transition occurs at approximately T = 200 K when the surface region becomes n-type (Figs. 1d,f), with the conduction band minimum below $E_F$. We examine thermo cycling measurements (Fig. 1h) and exclude potential aging effects for the band shift. Considering that the bulk carrier density only exhibits a negligible change in the region of 200 K < T < 300 K (Figs. 1b-c), the significant band shift observed by ARPES indicates a band bending near the surface region, as depicted in Fig. 1i.

Surprisingly, we observe a sensitive SPV at LILT. Figure 2a displays band structure near $E_F$ measured at T = 6 K, with the lowest incident light brightness of $1\times10^{-7}$ W/cm$^2$ (corresponding to $3\times10^{10}$ ph/(s·cm$^2$)) that we can use to collect ARPES spectra with reasonable statistics in an acceptable time (with intensity less than 1‰ of a typical lab-based He lamp brightness). We already observe considerable spectra weight above $E_F$, which can be better resolved in the energy distribution curve (EDC) in Fig. 2c. The spectra weight above $E_F$ is far beyond Fermi-edge broadening induced by temperature and energy resolution, as evidenced by the Fermi-edge of Cu reference. Furthermore, we observe a boost of spectra weight above $E_F$ with a notable upward shift of approximately 200 meV (Fig. 2b), as an ion gauge turned on that emits weak light in the measuring chamber (geometry illustrated in Fig. 2d). The direct comparison of EDCs with ion gauge on/off in Fig. 2c indicates a quasi-steady SPV at LILT, where considerable light-generated carriers in quasi-equilibrium cancel the surface band bending.

To quantitatively study the quasi-steady SPV, we performed incident light (hν = 21.2 eV) brightness-dependent measurements. The ion gauge is turned off, and other sources of illumination are excluded. As shown in Figs. 2e-f, we observe clear incident light-induced SPV, indicating that photogenerated holes during the photoemission process form a sufficient density in quasi-equilibrium and balance surface band bending. Simultaneously, the thermoelectric effect is excluded due to negligible thermal contributions at LILT. Figure 2g summarizes the incident light brightness-dependent shifts of leading edge midpoint (LEM) of conduction band (Fig. 2e) and valence band (Fig. 2f), that indicates a rigid band shift as expected from SPV in Fig. 2h. Notably, the corresponding optical responsivity achieves an impressive value of $2.4\times10^{6}$ V/(W·cm$^{-2}$) with light brightness of $3\times10^{-8}$ W/cm$^2$, significantly surpassing the previously record value of $4\times10^{4}$ V/(W·cm$^{-2}$) in HgCdTe [30].

Considering the narrow band gap of $NbSi_{0.5}Te_2$, we further evaluate the responsivity under infrared illumination. Based on a low He lamp brightness that exhibits weak SPV (Fig. 3a), additional continuous infrared illumination with wavelength of 980 nm is employed, with the results shown in Fig. 3b. We observe SPV effect with $\Delta V_{SPV} = 0.11$ eV with infrared illumination of $2.8\times10^{-5}$ W/cm$^2$ (Fig. 3b-I), yielding a infrared responsivity value of $1.8\times10^{3}$ V/(W·cm$^{-2}$). As infrared illumination increases up to $10^{-3}$ W/cm$^2$, SPV value reaches a maximum with $\Delta V_{SPV} = 0.16$ eV (Fig. 3b-II). We extract infrared brightness-dependent SPV values in Fig. 3c, which follow the same trend as that induced by incident light of hν = 21.2 eV.

By further increasing infrared illumination up to $2.5\times10^{-1}$ W/cm$^2$ (Figs. 3b-III to -V), SPV is suppressed and eventually approaches zero. This behavior is attributed to the Dember effect [31-33], which arises from the diffusion charge separation caused by the different mobilities of photogenerated electrons and holes (Fig. 3d). The Dember photovoltage is opposite to SPV in $NbSi_{0.5}Te_2$, leading to a reduction of $\Delta V_{SPV}$, whose non-monotonic behavior further provides conclusive evidence against thermoelectric contributions. The observation of the Dember effect indicates that the photogenerated carrier density is sufficiently high to build a notable diffusion current.

To understand the sensitive SPV observed in $NbSi_{0.5}Te_2$, we perform temperature-dependent measurements. We firstly focus on results with infrared illumination in strong brightness region of $8\times10^{-1}$ W/cm$^2$ (Fig. 4a). At T < 30 K, Dember effect dominates the system and completely cancels SPV due to the large photogenerated carrier density (Fig. 4c-I), consistent with the results in Fig. 3b. By increasing temperature above 30 K, the Dember effect weakens gradually and SPV reaches maximum 50 K (Fig. 4c-II). By further increasing temperature, SPV decreases slowly (Figs. 4c-III and IV). The extracted temperature-dependent SPV results are plotted in Fig. 4e. Remarkably, even at T = 160 K, which is significantly higher than liquid-nitrogen temperature, a notable quasi-steady SPV of approximately 60 meV persists in $NbSi_{0.5}Te_2$ (Fig. 4f). Considering that the photocurrent in the device has detectable responses even without a resolvable SPV established in quasi-equilibrium [34,35], our temperature-dependent results suggest that $NbSi_{0.5}Te_2$ has promising application potential in infrared detection.

As seen from Fig. 4e, the SPV results with strong illumination in high temperature region can be well described by the implicit expression of SPV [36],

$$\frac{\Delta V_{SPV}}{kT} e^{\frac{\Delta V_{SPV}}{kT}} = \delta e^{\frac{V_0}{kT}}, \tag{1}$$

where $V_0$ is the magnitude of the band bending, and $\delta = \Delta p / p_0$ is ratio between photogenerated carrier density $\Delta p$ and intrinsic bulk one $p_0$ in quasi-equilibrium.

In the low temperature region, the intrinsic bulk carrier density $p_0$ quickly drops and becomes negligible below 40 K, as seen from the transport data in Figs. 1b-c. Furthermore, intrinsic carrier freezing reduces the recombination rate and prolongs the lifetime of photogenerated carriers [37,38], which effectively increases photogenerated carrier density $\Delta p$ in quasi-equilibrium. Both effects together quickly boost $\delta$ and lead to the Dember effect at low temperature, which makes SPV deviation from the implicit formula (1).

Then we focus on temperature dependent results in the LILT region, with illumination of $3\times10^{-4}$ W/cm$^2$ from He-I$\alpha$ light (Fig. 4b). At T = 6 K, we observe a strong SPV effect (Fig. 4d-I), consistent with results in Fig. 2e. By rising temperature, SPV quickly drops and gets fully suppressed for T > 40 K (Fig. 4d). Such a strong temperature suppression of SPV in the LILT region cannot be explained by implicit SPV formula (1) (Supplementary Materials), due to the bulk carriers freezing. As discussed previously, the intrinsic bulk carrier density $p_0$ becomes extremely low for T < 40 K. Although the weak illumination builds a small amount of photogenerated carrier density $\Delta p$ in quasi-equilibrium, the freezing carriers can still induce a reasonable value of $\delta = \Delta p / p_0$, which leads to strong quasi-steady SPV in the LILT region. Intrinsic carrier density $p_0$ gradually increases as temperature rises, resulting in a quick drop of $\delta$ and the suppression of $\Delta V_{SPV}$. We note that the emergence of Dember/SPV effect in the high/low illumination condition (Fig. 4a/b) shares a similar temperature window, due to the same mechanism of bulk carriers freezing.

In summary, we demonstrate a sensitive quasi-steady SPV in a layered narrow gap semiconductor $NbSi_{0.5}Te_2$, achieving a remarkable photoresponsivity of $2.4\times10^6$ V/(W·cm$^{-2}$) at LILT. As infrared illumination exceeds $10^{-3}$ W/cm$^2$, we directly observe the Dember effect, where the photogenerated carrier density in quasi-equilibrium is high enough to drive diffusion current and cancel SPV. Further temperature-dependent measurements indicate that low-temperature bulk carriers freezing results in the ultrahigh SPV efficiency at LILT and the Dember effect. Given the potential to enhance optical responsivity via device fabrication, this layered semiconductor $NbSi_{0.5}Te_2$ shows significant promise for practical applications in detecting low-intensity infrared signals. Its ultrahigh photoresponse at LILT makes this material especially suitable for advanced optoelectronic sensing technologies, including deep space exploration, polar research, remote sensing, and space communication. Moreover, our results provide another case of observer effect [39], that ARPES measurement process itself, even when the incident light is much weaker than usually used, can drive the system deviated from the original state, with considerable photogenerated carrier density in quasi-equilibrium.

## Methods

**Sample synthesis.** Single crystals of $NbSi_{0.5}Te_2$ were successfully synthesized by the flux method. The raw materials Nb (slug, 99.95%, Alfa Aesar), Si (lump, 99.99%, Alfa Aesar), and Te (shot, 99.999%, Tangchuan Sci-Tech Co., Ltd) were mixed with a mole ratio of 1:3:8 and placed into an alumina crucible. The crucible was then sealed in a quartz tube under high-vacuum conditions. The tube was heated to 1423 K over 15 h and maintained for 20 h. It was then slowly cooled to 1123 K at a rate of 2 K/h. Single crystals of $NbSi_{0.5}Te_2$ were obtained by immediately centrifuging the excess Si-Te flux.

**Excitation light.** The excitation light of 21.2 eV is generated by a He lamp (He Iα), and its intensity is adjusted via an aluminum window and a vacuum valve. The infrared excitation light is produced by a continuous infrared laser with a power of 630 mW, whose intensity is modulated by optical filters possessing varying transmittance properties.

**ARPES**. Clean surfaces for ARPES measurements were obtained by in situ cleaving of samples in a vacuum chamber maintained at a pressure below $5 \times 10^{-11}$ Torr. The main ARPES measurements were conducted at a lab-based ARPES, utilizing a He lamp as the photon source (He Iα: 21.2 eV). The photon-energy-dependent ARPES measurements were conducted at Dreamline of the Shanghai Synchrotron Radiation Facility.

**ACKNOWLEDGMENTS**

We thank the support with the ARPES experiments by BL03U and BL07U beamlines of the Shanghai Synchrotron Radiation Facility, and BL13U beamline of the National Synchrotron Radiation Laboratory. This work was supported by the National Natural Science Foundation of China (NSFC) (Grants No. 12274329, and No. 12404083), New Cornerstone Science Foundation (Grant No. 23H010801236), Innovation Program for Quantum Science and Technology (Grant No. 2021ZD0302700), National Key Research and Development Program of China (Grant No. 2024YFA140840), the China Postdoctoral Science Foundation (Grant No. 2023M732717), and the SynergeticExtreme Condition User Facility (SECUF).

[1] R. H. Hadfield, *Single-photon detectors for optical quantum information applications.* Nat. Photonics **3**, 696 (2009).

[2] J. A. Lau, V. B. Verma, D. Schwarzer and A. M. Wodtke, *Superconducting single-photon detectors in the mid-infrared for physical chemistry and spectroscopy*, Chem. Soc. Rev. **52**, 921 (2023).

[3] I. Kviatkovsky, H. M. Chrzanowski, E. G. Avery, H. Bartolomaeus, S, Ramelow, *Microscopy with undetected photons in the mid-infrared*, *Sci. adv.* **6** eabd0264 (2020).

[4] Y. N. Han, A. K. Zhang, *Cryogenic technology for infrared detection in space*, Sci. Rep. **12**, 2349 (2022).

[5] L. Kronik, Y. Shapira, *Surface photovoltage phenomena: theory, experiment, and applications*. Surf. Sci. Rep. **37**, 1 (1999).

[6] L. Kronik, Y. Shapira, *Surface photovoltage spectroscopy of semiconductor structures: at the crossroads of physics, chemistry and electrical engineering*. Surf. Interface Anal. **31**, 954 (2001).

[7] D. K. Schroder, *Surface voltage and surface photovoltage: history, theory and applications*. Meas. Sci. Technol. **12**, R16 (2001).

[8] C. -H. Ho, Y. -J. Chu, *Bending Photoluminescence and Surface Photovoltaic Effect on Multilayer InSe 2D Microplate Crystals*. Adv. Optical Mater. **3**, 1750 (2015).

[9] S. Fang, D. H. Wang, Y. Kang, X. Liu, Y. M. Luo, K. Liang, L. N. Li, H. B. Yu, H. C. Zhang, M. H. Memon, B. Y. Liu, Z. H. Liu, and H. D. Sun, *Balancing the Photo-Induced Carrier Transport Behavior at Two Semiconductor Interfaces for Dual-Polarity Photodetection*, Adv. Funct. Mater. **32**, 2202524 (2022).

[10] L. Wang, I. Meric, P. Y. Huang, Q. Gao, Y. Gao, H. Tran, T. Taniguchi, K. Watanabe, L. M. Campos, D. A. Muller, J. Guo, P. Kim, J. Hone, K. L. Shepard, and C. R. Dean, *One-dimensional electrical contact to a two-dimensional material*. Science **342**, 614 (2013).

[11] Q. H. Wang, K. Kalantar-Zadeh, A. Kis, J. N. Coleman, and M. S. Strano, *Electronics and optoelectronics of two-dimensional transition metal dichalcogenides*. Nat. Nanotechnol. **7**, 699 (2012).

[12] A. Avsar, H. Ochoa, F. Guinea, B. Özyilmaz, B. J. van Wees, and I. J. Vera-Marun, *Colloquium: Spintronics in graphene and other two-dimensional materials.* Rev. Mod. Phys. **92**, 021003 (2020).

[13] H. Kurebayashi, J. H. Garcia, S. Khan, J. Sinova, and S. Roche, *Magnetism, symmetry and spin transport in van der Waals layered systems*. Nat. Rev. Phys., **4**, 150 (2022).

[14] C. Liu, H. Chen, S. Wang, Q. Liu, Y. G. Jiang, D. W. Zhang, M. Liu, and P. Zhou, *Two-Dimensional Materials for Next-Generation Computing Technologies*. Nat. Nanotechnol. **15**, 545 (2020).

[15] S. M. Zhang, X. N. Deng, Y. F. Wu, Y. Q. Wang, S. X. Ke, S. S. Zhang, K. Liu, R. T. Lv, Z. C. Li, Q. H. Xiong, and C. Wang, *Lateral layered semiconductor multijunctions for novel electronic devices*. Chem. Soc. Rev. 51, 4000 (2022).

[16] Q. H. Wang, K. K.-Zadeh, A. Kis, J. N. Coleman, and M. S. Strano, *Electronics and optoelectronics of two-dimensional transition metal dichalcogenides*. Nat. Nanotechnol. **7**, 699 (2012).

[17] X. L. Chen, X. B. Lu, B. C. Deng, O. Sinai, Y. C. Shao, C. Li, S. F. Yuan, V. Tran, K. Watanabe, T. Taniguchi, D. Naveh, L. Yang, and F. N. Xia, *Widely tunable black phosphorus mid-infrared photodetector*, Nat. Commun. **8**, 1672 (2017).

[18] H. T Yuan, X. G. Liu, F. Afshinmanesh, W. Li, G. Xu, J. Sun, B. Lian, A. G. Curto, G. J. Ye, Y. Hikita, Z. X. Shen, S. -C. Zhang, X. H. Chen, M. Brongersma, H. Y. Hwang, and Y. Cui, *Polarization-sensitive broadband photodetector using a black phosphorus vertical p–n junction*. Nat. Nanotech. **15**, 675 (2020).

[19] Z. S. Chen, J. W. Dong, C. Giorgetti, E. Papalazarou, M. Marsi, Z. L. Zhang, B. B. Tian, Q. W. Ma, Y. C. Cheng, J. -P. Rueff, A. T.-Ibrahimi, and L. Perfetti, S*pectroscopy of buried states in black phosphorus with surface doping*. 2D Mater. **7**, 035027 (2020).

[20] G. Kremer, M. Rumo, C. Yue, A. Pulkkinen, C. W. Nicholson, T. Jaouen, F. O. von Rohr, P. Werner, and C. Monney, *Ultrafast dynamics of the surface photovoltage in potassium-doped black phosphorus*. Phys. Rev. B **104**, 035125 (2021).

[21] T. Y. Yang, Q. Wan, D. Y. Yan, Z. Zhu, Z. W. Wang, C. Peng, Y. B. Huang, R. Yu, J. Hu, Z. Q. Mao, S. Li, S. Y. A. Yang, H. Zheng, J. F. Jia, Y. G. Shi, and N. Xu, *Directional massless Dirac fermions in a layered van der Waals material with onedimensional long-range order*. Nat. Mater. **19**, 27 (2020).

[22] S. Li, Y Liu, S. S. Wang, Z. M. Yu, S. Guan, X. L. Sheng, Y. G. Yao, and S. Y. A. Yang, *Nonsymmorphic-symmetry-protected hourglass Dirac loop, nodal line, and Dirac point in bulk and monolayer* $X_3SiTe_6$ *(X = Ta, Nb)*, Phys. Rev. B 97, 045131 (2018).

[23] T. Sato, Z. W. Wang, K. Nakayama, S. Souma, D. Takane, Y. Nakata, H. Iwasawa, C. Cacho, T. Kim, T. Takahashi, and Y. Ando, *Observation of band crossings protected*

*by nonsymmorphic symmetry in the layered ternary telluride $Ta_3SiTe_6$*, Phys. Rev. B 98, 121111(R) (2018).

[24] Q. Wan, T. Y. Yang, S. Li, M. Yang, Z. Zhu, C. L. Wu, C. Peng, S. K. Mo, W. Wu, Z. H. Chen, Y. B. Huang, L. L. Lev, V. N. Strocov, J. Hu, Z. Q. Mao, Hao Zheng, J. F. Jia, Y. G. Shi, Shengyuan A. Yang, and N. Xu, *Inherited weak topological insulator signatures in the topological hourglass semimetal $Nb_3XTe_6$ (X = Si, Ge)*. Phys. Rev. B **103**, 165107 (2021).

[25] F. Boucher, V. Zhukov, and M. Evain, *$MA_xTe_2$ Phases (M = Nb, Ta; A = Si, Ge; 1/3 ⩽ x ≤ 1/2): An Electronic Band Structure Calculation Analysis.* Inorg. Chem. **35**, 7649 (1996).

[26] J. Zhang, Z. l. Yang, S. Liu, W. Xia, T. S. Zhu, C. Chen, C. W. Wang, M. X. Wang, S. -K. Mo, L. X. Yang, X. F. Kou, Y. F. Guo, H. J. Zhang, Z. K. Liu, and Y. L. Chen, *Direct Visualization and Manipulation of Tunable Quantum Well State in Semiconducting $Nb_2SiTe_4$.* ACS. Nano **15**, 15850 (2021).

[27] M. X. Zhao, W. Xia, Y. Wang, M. Luo, Z. Tian, Y. F. Guo, W. D. Hu, and J. M. Xue, *$Nb_2SiTe_4$: A Stable Narrow-Gap Two-Dimensional Material with Ambipolar Transport and Mid-Infrared Response*. ACS Nano **13**, 10705 (2019).

[28] K. Y. Zhou, J. Deng, L. Chen, W. Xia, Y. F. Guo, Y. Yang, J. -G. Guo, and L. W. Guo, *Observation of large in-plane anisotropic transport in van der Waals semiconductor $Nb_2SiTe_4$*, Chin. Phys. B **30**, 8 (2021).

[29] B. B. Wang, W. Xia, S. Li, K. Wang, S. Y. A. Yang, Y. F. Guo, and J. M. Xue, *One-Dimensional Metal Embedded in Two-Dimensional Semiconductor in $Nb_2Si_{x-1}Te_4$*. ACS. Nano **15**, 7149 (2021).

[30] G. P. Eppeldauer, R. J. Martin, *Photocurrent Measurement of PC and PV HgCdTe Detectors*, J. Res. Natl. Inst. Stand. Technol. **106**, 577 (2001).

[31] H. Dember, *A photoelectrical-motor energy in copper-oxide crystals*. Phys. Z. **32**, 554 (1931).

[32] S. R. Goldman, K. Kalikstein, and B. Kramer, Dember-effect theory. J. Appl. Phys. **49**, 2849 (1978).

[33] R. T. Chen, S. Pang, H. Y. An, J. Zhu, S. Ye, Y. Y. Gao, F. T. Fan, and C. Li, *Charge separation via asymmetric illumination in photocatalytic $Cu_2O$ particles*. Nat. Energy **3**, 655 (2018).

[34] F. Liu, C. Zhu, L. You, S.J. Liang, S. Zheng, J. Zhou, Q. Fu, Y. He, Q. Zeng, H. J. Fan, L.K. Ang, J. Wang, and Z. Liu, *2D Black Phosphorus/$SrTiO_3$-Based Programmable Photoconductive Switch*. Adv. Mater. **28**, 7768 (2016).

[35] C. Chen, N. Youngblood, R. Peng, D. Yoo, D. A. Mohr, T.W. Johnson, S. H. Oh, and M. Li, *Three-Dimensional Integration of Black Phosphorus Photodetector with Silicon Photonics and Nanoplasmonics*. Nano Lett. **17**, 985 (2017).

[36] J. P. Long, H. R. Sadeghi, J. C. Rife, and M. N. Kabler, *Surface space-Charge dynamics and surface recombination on silicon (111) surfaces measured with combined laser and synchrotron radiation*. Phys. Rev. Lett. **64**, 1158 (1989).

[37] T. Kohno, Y. Sudo, M. Yamauchi, K. Mitsui, H. Kudo, H. Okagawa, and Y. Yamada, *Internal Quantum Efficiency and Nonradiative Recombination Rate in InGaN-Based Near-Ultraviolet Light-Emitting Diodes*. Jpn. J. Appl. Phys. **51**, 072102 (2012).

[38] Z. C. Su, S. J. Xu, *Effective lifetimes of minority carriers in time-resolved photocurrent and photoluminescence of a doped semiconductor: Modelling of a GaInP solar cell.* Sol Energy Mater Sol Cells **193**, 292 (2019).

[39] A. Zeilinger, *Experiment and the foundations of quantum physics*. Rev. Mod. Phys. **71**, S288 (1999).

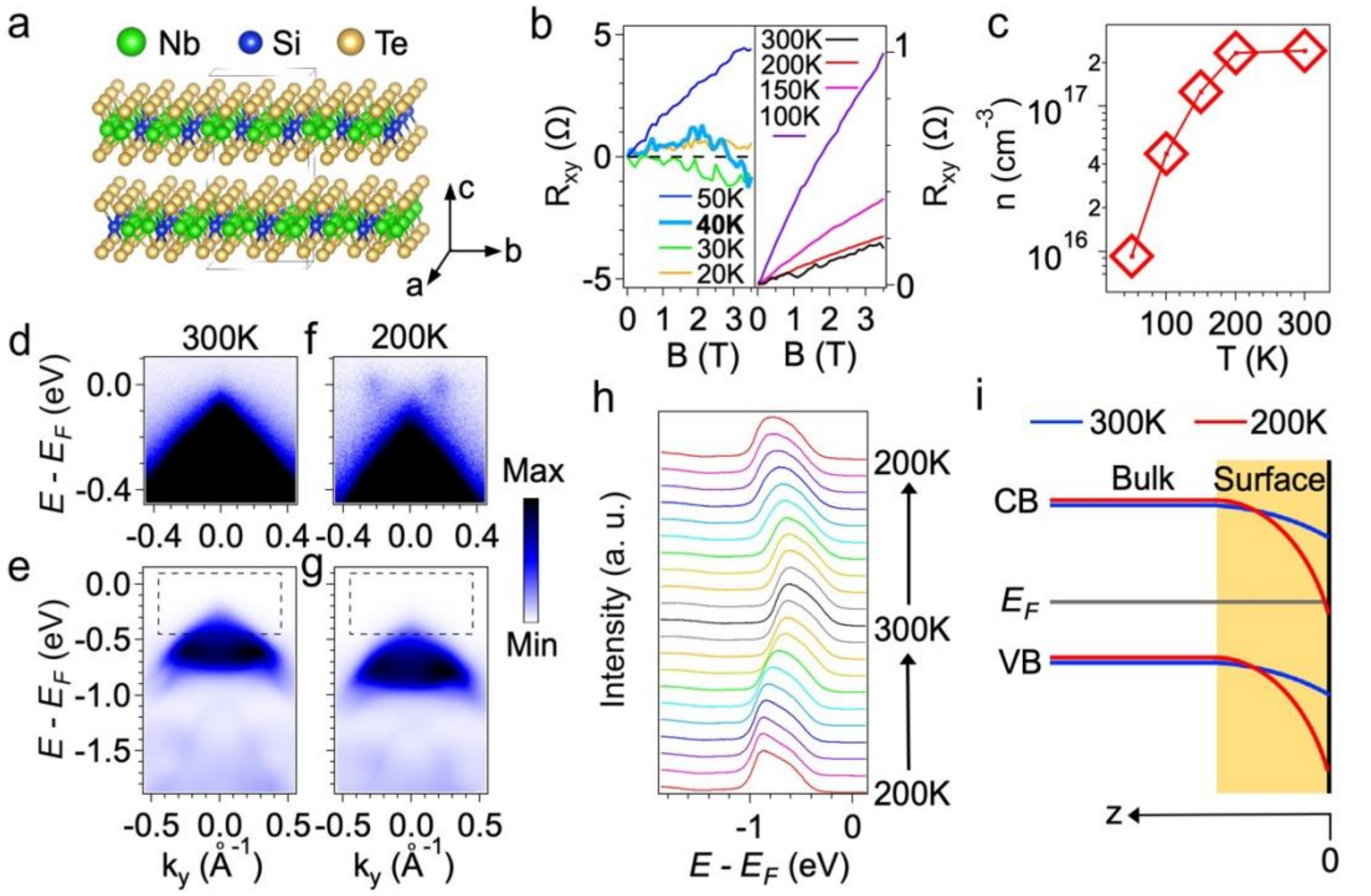

**Figure 1. Surface band bending in $NbSi_{0.5}Te_2$. a,** The crystal structure of $NbSi_{0.5}Te_2$. **b,** Temperature-dependent Hall resistivity. **c,** Corresponding carrier density as a function of temperature. **d-e,** Photoemission intensity plots at T = 300 K. **f-g,** Same as **d-e**, but with T = 200 K. **h,** Temperature-dependent EDCs at the Γ point. **i,** Schematic diagram of surface band bending.

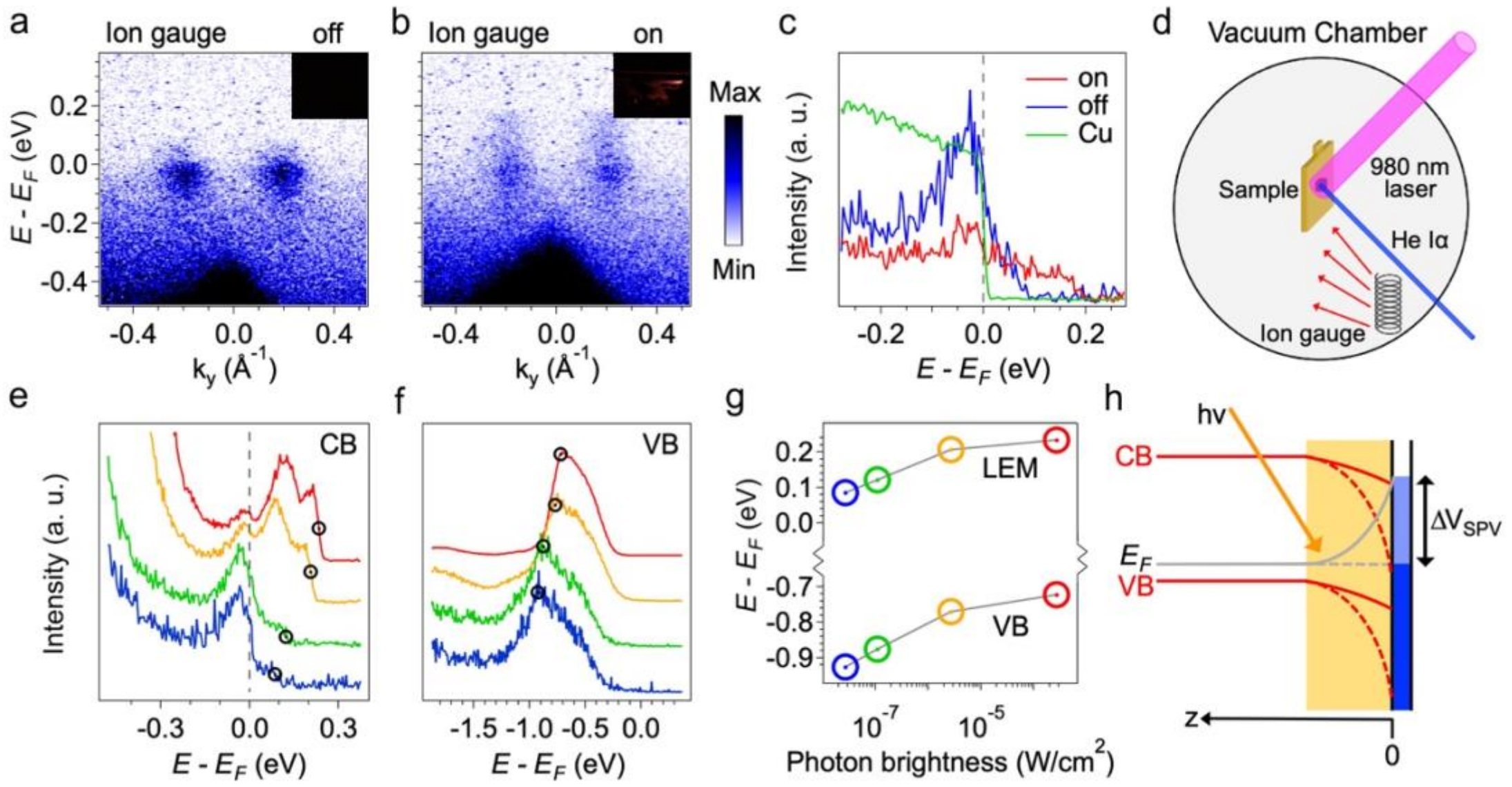

**Figure 2. Incident He I$\alpha$ light induced SPV at T = 6 K. a-b,** Photoemission intensity plots with incident He I$\alpha$ light brightness of $1\times10^{-7}$ W/cm$^2$, with ion gauge on and off, respectively. Inset: photos of the sample with ion gauge on/off. **c,** Corresponding EDCs at the conduction band minimum. The result on Cu is appended for reference. **d,** The geometric arrangement of the ARPES experiment. **e-f,** Brightness-dependent EDC at conduction and valence bands, respectively. The blue, green, yellow and red lines correspond to $3\times10^{-8}$, $1\times10^{-7}$, $3\times10^{-6}$, and $3\times10^{-4}$ W/cm$^2$. **g,** The extracted brightness-dependent leading edge midpoint (LEM) and valence band position, with the LEM matching the Fermi level $E_F$. **h,** Schematic diagram of SPV.

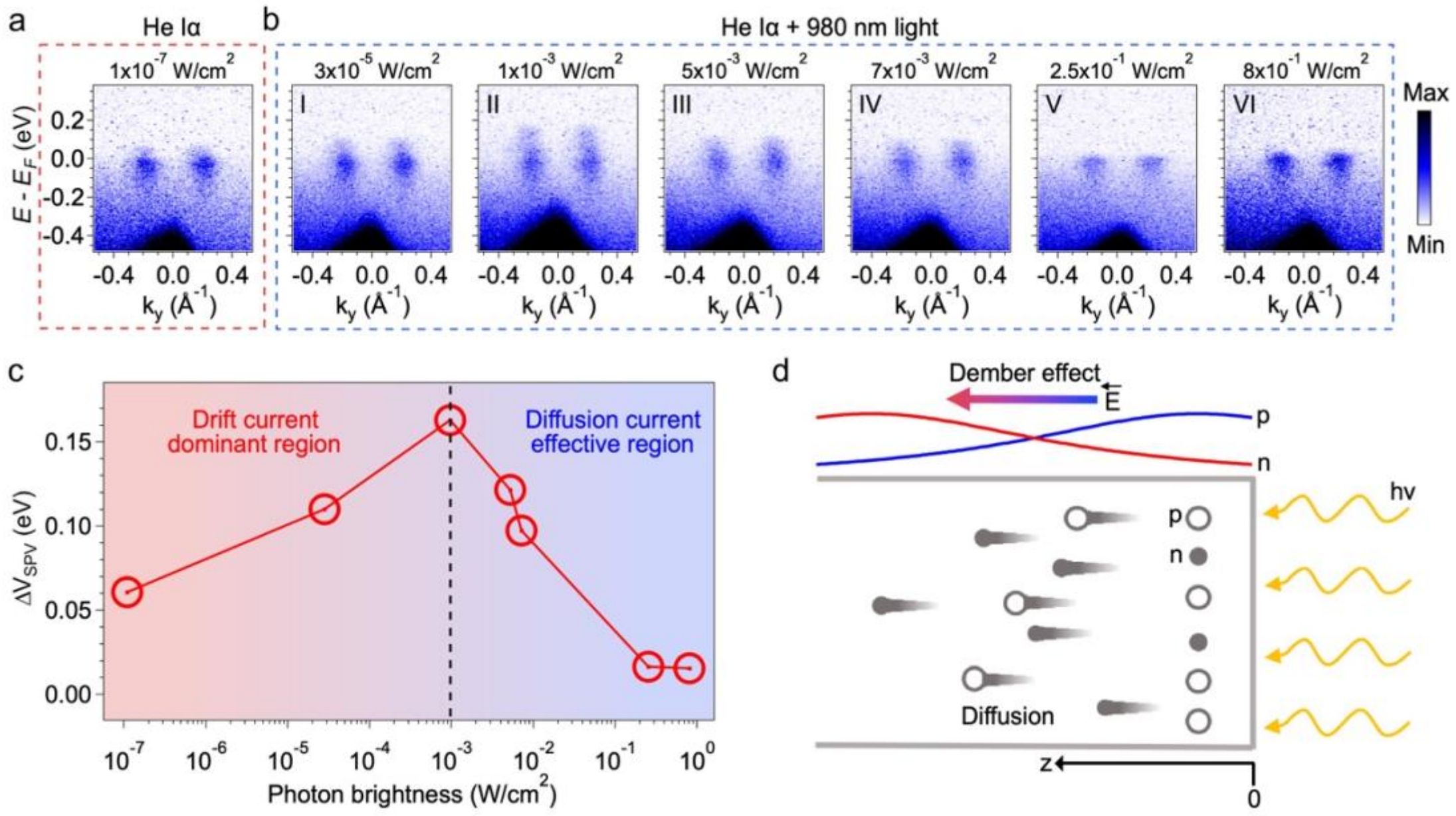

**Figure 3. SPV in infrared illumination region T = 6 K . a,** Photoemission intensity plots with incident He I$\alpha$ light brightness of $1\times10^{-7}$ W/cm$^2$, with ion gauge off. **b,** Same as a, with 980 nm infrared illumination of different brightness. **c**, Extracted Brightness-dependent SPV value $\Delta V_{SPV}$. **d,** Schematic diagram of Dember effect.

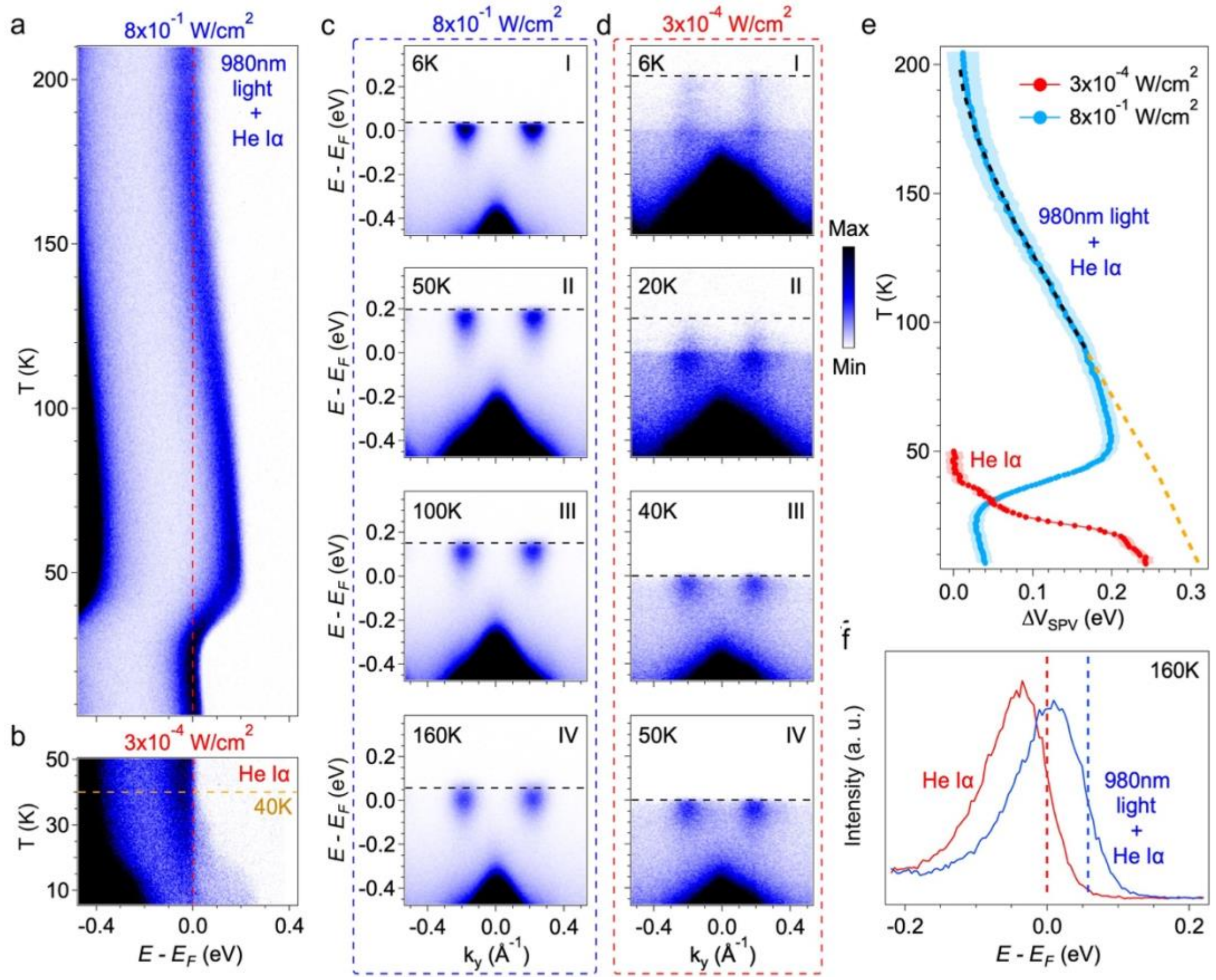

**Figure 4. Temperature-dependent SPV. a,** Temperature-dependent photoemission data at conduction band, with 980 nm infrared illumination brightness of 8×10$^{-1}$ W/cm$^2$. **b,** Same as **a**, but with He I$\alpha$ light brightness of $3 \times 10^{-4}$ W/cm$^2$. **c**, Corresponding photoemission intensity plots of **a** at T = 6 K, 50 K, 100 K and 160 K. **d**, Corresponding photoemission intensity plots of **b** at T = 6 K, 20 K, 40 K and 50 K. **e,** Temperature-dependent SPV value $\Delta V_{SPV}$ extracted from **a** and **b**, the shaded areas correspond to the error bars, the dashed black line represents the theoretically simulated temperature-dependent SPV value given by Eq. (1). **f,** The EDCs at the conduction band with (blue line) and without (red line) infrared illumination of 8×10$^{-1}$ W/cm$^2$ at T = 160 K.